\documentclass[12pt]{article}
\usepackage{amsmath,amssymb,amsthm}
\usepackage{natbib}
\usepackage{booktabs}
\usepackage{enumerate}

\newtheorem{assump}{Assumption}
\newtheorem{thm}{Theorem}
\newtheorem{lemma}{Lemma}  
\newtheorem{example}{Example} 

\newcommand{\E}{\mathbb{E}}
\newcommand{\Var}{\operatorname{Var}}
\newcommand{\ATT}{\mathrm{ATT}}

\DeclareMathOperator*{\argmin}{arg\,min}  

\title{Consistency of an Intercept-Shifted Synthetic-Control Estimator under Weighted Parallel Trends}
\author{Michael Guggisberg\thanks{Cisco Systems, Los Angeles, California, USA. Email: mikeguggis@gmail.com.}}
\date{\today}

\begin{document}
	
	\maketitle
	
	\begin{abstract}
		The average treatment effect on the treated (ATT) in a staggered-adoption panel is estimated using an intercept-augmented synthetic-control (SCM) estimator.
		A \emph{weighted parallel trends} plus an intercept shift,
		together with mild regularity on the weight vectors
		(non-degenerate dispersion) and expanding pre-treatment length,
		are sufficient for consistency allowing for heavy-tailed shocks.  These conditions can be more
		interpretable than the autoregressive or low-rank factor models with light tails
		assumed by Ben-Michael, Feller, and Rothstein (2022) and expand the valid DGP pool from the same paper.  Practical diagnostics
		to support the assumptions are discussed and situate these results within the recent
		literature on SC + DiD hybrids.
	\end{abstract}
	
	\section{Introduction}
	The synthetic-control method
	\citep{abadie2010synthetic, abadie2015comparative}
	and the multi-period difference-in-differences (DiD) family
	\citep{callaway2021difference,sun2021estimating,
		athey2021design} are now the work-horses for policy evaluation with
	staggered adoption.
	Ben-Michael, Feller, and Rothstein ``Synthetic controls with staggered adoption" (2022, henceforth \textit{SCSA}) \nocite{benmichael2022synthetic}
	extended SCM to multiple treated units under a staggered adoption scheme and derived finite-sample error
	bounds under (i)~a time-varying autoregressive law or (ii)~a linear
	factor structure.  While powerful, these structural assumptions can be
	difficult to motivate in many applications.
	
	One can retain the attractive \emph{weighting} logic of
	SCM while identifying ATT under a simpler, DiD-style assumption.
	The key ingredient is to add an \emph{intercept shift} proposed by \textit{SCSA}, called ``augmented SCM'' (or intercept-augmented SCM).  It is shown that
	(1) Stable Unit Treatment Value Assumption (SUTVA) and (2)~weighted parallel trends are sufficient for consistency without requiring light-tail errors (provided the weights do not degenerate onto a single donor).  The resulting
	conditions can be assessed with the usual placebo-gap plots, weight
	summaries and window-length sensitivity, and sidestep the need to
	argue for AR($L$) or factor models.
	
	Two papers point in the same direction:
	\citet{doudchenko2016difference} blend SCM with DiD via an intercept and
	\citet{arkhangelsky2021synthetic} introduce ``synthetic DiD'' in a
	simultaneous-adoption setting. 	To author's knowledge, no published work provides (i) a finite-sample error
	bound for the staggered-adoption ATT that separates pooled and
	unit-specific imbalance (Theorem \ref{thm:fsp}) under this alternative assumption set, (ii) general conditions for consistency of weighted parallel trends (Theorem \ref{thm:consistency}), nor (iii) an arg-min consistency
	result for the partially-pooled ridge-regularised estimator used in \textit{SCSA} (Theorem \ref{thm:scsa_consistency}); this paper fills
	that gap.

	\section{Setup and notation}
	Units are indexed \(i=1,\dots,N\); time periods are
	\(t=1,\dots,T\).  Treatment arrives once and is irreversible.
	Let \(T_i\in\{1,\dots,T,\infty\}\) be unit \(i\)'s adoption date, where $T_i = \infty$ are untreated. The population and population size of treated units is 
	\(\mathcal J=\{i:T_i<\infty\}\) and \(J=|\mathcal J|\). 
	Potential outcomes at period $t$ are \(Y_{it}(s)\) for a scenario in which \(i\)
	adopts at \(s\); by convention \(Y_{it}(\infty)\equiv Y_{it}^0\).
	
	\paragraph{Observed data and estimand.}
	The \(Y_{it}=Y_{it}(T_i)\) is observed.
	Define the \emph{population} ATT at event time \(k\ge0\):
	\begin{align*}
		\tau_{jk} \;&=\;Y_{j,T_j+k}(T_j)-Y_{j,T_j+k}(\infty) \\
		\ATT_k \;&=\;
	\frac1J\sum_{j\in\mathcal J}
	\E[\tau_{jk}]\\
	\bar{\ATT} \;&=\; \frac{1}{K}\sum_{k=1}^{K} \ATT_k.
\end{align*}
	
	\paragraph{Estimator.}
	For each treated unit \(j\) choose non-negative weights
	\(w_{ij}\) where $\sum_i w_{ij}=1$ and satisfying the donor restriction
	\(w_{ij}=0\) if \(i=j\) or \(T_i\le T_j\).  Let
	\(L_i=T_i-1\ge1\) be the available pre-treatment length and set
	\(\bar Y_{i,\text{pre}}=L_i^{-1}\sum_{s=1}^{T_i-1}Y_{is}\).
	Define demeaned series \(Z_{it}=Y_{it}-\bar Y_{i,\text{pre}}\).
	The intercept-shifted SCM estimator is
	\begin{align*}
	\hat\tau_{jk}\;&=\;	Z_{j,T_j+k}-\sum_{i}w_{ij}Z_{i,T_j+k}, \\
	\widehat{\ATT}_k&=\frac1J\sum_{j\in\mathcal J}\hat\tau_{jk}, \\
	\widehat{\bar{\ATT}} &= \frac{1}{K}\sum_{k=1}^{K} \widehat{\ATT}_k.
	\end{align*}
	
	\section{Assumptions}
	
	Below are the assumptions used for the main results. Assumptions~\ref{assump:anticip}–\ref{assump:subE} are used in Theorems \ref{thm:fsp} and \ref{thm:consistency}. All assumptions (\ref{assump:anticip}-\ref{assump:C2}) are used in Theorem \ref{thm:scsa_consistency}.
	
	Assumptions~\ref{assump:anticip}–\ref{assump:sutva} amount to the SUTVA in
	the staggered-adoption environment.

	\begin{assump}[No anticipation]\label{assump:anticip}
		For every \(i\) and every \(s>t\),
		\(Y_{it}(s)=Y_{it}(\infty)\).
	\end{assump}
	
	\begin{assump}[No interference]\label{assump:sutva}
		Let \(\mathbf s=(s_1,\dots,s_N)\) and
		\(\mathbf s'=(s'_1,\dots,s'_N)\) be two adoption calendars.
		If \(s_i=s'_i\) then \(Y_{it}(s_i)=Y_{it}(\mathbf s)=Y_{it}(\mathbf s')\) for all
		\(t\).
	\end{assump}

	Assumption~\ref{assump:wpt} is a weighted version of the usual parallel-trends hypothesis. It replaces the structural restrictions	imposted in \textit{SCSA}, time-varying ARL(L) (their Assumption 2a) or a low‑rank factor structure (2b)—with a design‑type restriction that is already central to the DiD literature.  The idea is to search—not over parametric models—but over weight vectors that make the (possibly heterogeneous) donor tend ``look like" unit $j$.
	
	\begin{assump}[Weighted parallel trends]\label{assump:wpt}
		For each treated unit \(j\) there exists
		\((w_{1j},\dots ,w_{Nj})\) such that for
		every \(t<T_j\)
		\[
		\sum_i w_{ij}\,
		\E\!\bigl[\Delta Y_{it}^0\bigr]
		\;=\;
		\E\!\bigl[\Delta Y_{jt}^0\bigr],
		\]
		where \(\Delta Y_{it}^0 := Y_{it}^0-Y_{i,t-1}^0\).
	\end{assump}
	
	Assumption \ref{assump:wpt} implicitly requires that the donor pool be rich enough to find weights that replicate the treated units’ pre-treatment trajectory. The moment conditions can only be satisfied if the stacked vector of lagged outcomes for the treated units lies (approximately) in the convex hull of the corresponding vectors for the donors. If the donor pool is too small, or if its members follow very different persistent trends, such weights need not exist and the assumption would fail. Assumption \ref{assump:C1} formalizes this by requiring that convex-hull/spanning of the lagged-outcome space is sufficient for (near-exact) weights to exist as either the number of donors or the length of the pre-period grows. Thus, Assumption \ref{assump:wpt} should be read as both a parallel-trends restriction and an existence statement whose plausibility rises with the richness of the donor pool.
	
	 The weighted parallel trends assumption is implied by \textit{SCSA} Assumptions 2a and 2b. A weighted parallel trends assumption is also used in Synthetic Difference-in-Differences and inclusive-SCM \citep{inclusiveSCM2024,arkhangelsky2021synthetic}.

	Assumption~\ref{assump:weights} prevents
	the weights from degenerating on a single donor; ridge or entropy
	penalties used in empirical SCM help enforce this. 
	
	\begin{assump}[Weight regularity]\label{assump:weights}
(a) Donor-restriction: for each treated unit \(j\) and every event time $k\geq 0$, the weight vector \(w_j\) lies in the simplex and satisfy $w_{ij}=0$ whenever $T_i \leq T_j + k$. (b) Uniform non-degeneracy: there exists a constant \(c\in(0,1)\) such that
\[
\sum_{i} w_{ij}^{2}\;\le\; c \quad\text{for every treated unit }j .
\]
	\end{assump}
	
	While the SCSA DGP assumptions satisfy weighted parallel trends, \textit{SCSA} Assumption 2b can violate the combination of Assumptions \ref{assump:wpt}-\ref{assump:weights} without an additional restriction -- such as equality of the group-average loadings. However, if weights were unrestricted as in \cite{doudchenko2016difference}, then \textit{SCSA} DGPs would be fully captured without restriction. 
	
	Additionally, this assumption allows the donor pool to change with $k$. \textit{SCSA} use a slightly stronger condition where there is zero weight for donors treated at or before unit $j$. Although, they mention their estimator does not require this restriction.
	
	Let $N_{0}:=\max_{j,k}|\{i:T_{i}>T_{j}+k\}|$ denote the maximum number of units that ever qualify as donors.  Further define $L_{\min}:=\min_{i}L_i$ and  $L_{\max}:=\max_{i}L_i$  to be the shortest and longest pre-treatment window in the sample;  
	note that \(L_{\min}\le L_{\max}\le T\).  Assumption~\ref{assump:L} is the asymptotic scheme stating that i) pre-treament periods is increasing or the treatment periods shift later as the panel lengthens, ii) the effective dimension of the weight simplex is controlled, and iii) number of treated units is increasing. 
	
	\begin{assump}[Asymptotic scheme]\label{assump:L}
		As $(N,T)\to\infty$,
			\begin{enumerate}[(i)]
			\item $L_{\min}\;\to\;\infty,$
			\item $(N_{0}-1)J=o\!\bigl(L_{\max}/\log L_{\max}\bigr),$ \text{and}
			\item $\max\!\Bigl\{\tfrac{\log J}{J},\,\tfrac1{J\,K}\Bigr\}\;\to\;0.$
		\end{enumerate}
	\end{assump}
	
	Note, all simplices that the
	estimator searches over are embedded in $\mathbb R^{N_{0}}$, so the covering
	number in Lemma~\ref{lem:ULLN} is valid for every $(j,k)$.
	
	Assumption~\ref{assump:subE} controls tail behavior and allows heavier tails (e.g., Gumbel distribution) than the typically assumed sub-Gaussian tails.

	\begin{assump}[Sub-exponential shocks]\label{assump:subE}
		Write \(Y_{it}^0=\mu_{it}+\varepsilon_{it}\) with
		\(\E[\varepsilon_{it}]=0\); the \(\varepsilon_{it}\) are independent
		across \(i,t\) and are sub-exponential with scale parameter \(\kappa<\infty\).
 The same $\varepsilon_{it}$ enters every potential outcome
		$Y_{it}(s)$, $s\in\{1,\dots,T,\infty\}$.
	\end{assump}
	
Note that the assumption can be trivially weaked to allow $(i,t)$ pairs to have their own scale parameters. However, they would still be considered sub-exponential to the largest scale parameter.

Assumption~\ref{assump:C1} states the donor pool is “rich” enough that, as
either the number of never-treated units or the pre-period length
grows, one can find weights that achieve arbitrarily close
	balance on pre-treatment outcomes.  This occus when (i) Under the AR($L$) or low-rank
factor models used in \textit{SCSA};
(ii) when donors span the treated units’ lagged-outcome convex
hull.

	\begin{assump}[Convex-hull / spanning]\label{assump:C1}
		For every sample size $(N,T)$ there exists a weight matrix
		$\Gamma^\star\!\in\!\Delta_{\!scm}$ satisfying
		\[
		q_{\mathrm{pool}}(\Gamma^\star) \,\le\, \eta_{N,T},
		\qquad
		q_{\mathrm{sep}}(\Gamma^\star)  \,\le\, \eta_{N,T},
		\quad\text{with } \eta_{N,T}\downarrow0
		\text{ as } N_0\text{ or }L_{\min}\to\infty .
		\]
	\end{assump}
	
	The terms $q_{\mathrm{pool}}$ and $q_{\mathrm{sep}}$ are defined in the proof of Theorem \ref{thm:fsp}. The term $q_{\mathrm{pool}}$ captures systematic, group-level mismatch by pooling all treated units for each lag $\ell$, taking the donor-weighted gap, then square and average over lags. While $q_{\mathrm{sep}}$ captures the idiosyncratic, unit-specific mismatch by obtaining a squared donor gap for each treated unit, averaging over lags, then averaging over units.
	
	This assumption can fail when a treated unit with trends or seasonality that
	never appears in the donors; a donor pool that is too small or a donor pool whose outcomes are so sparsely distributed that, even after de-meaning, no convex combination can track the treated path.
	This assumption can be checked by plotting RMSE placebo gaps versus $L$ or versus donor-pool size. A monotone decline toward zero supports the assumption.  
	
Assumption~\ref{assump:C2} states the penalty becomes weak enough that it
does not block the optimiser from
approaching the best attainable fit (order $\eta_{N,T}^2$),
yet strong enough to keep the weight norm
$\Vert\hat\Gamma\Vert_F$ bounded.

	\begin{assump}[Regularisation tuning]\label{assump:C2}
		The ridge parameter in the partially-pooled objective satisfies
		$\lambda=\lambda_{N,T}\downarrow0$ and
		$\lambda^{-1}\eta_{N,T}^2\to0$ as $(N,T)\to\infty$.
	\end{assump}
	
	This assumption can fail when choosing a fixed, large \(\lambda\)
	independent of sample size (violating the first condition) or
	setting \(\lambda=0\) making weights degenerate (violating the second condition).
	This assumption can be checked by grid-search \(\lambda\) against placebo-gap
	cross-validation; make sure the selected value declines when additional donors or pre-periods are included.

	\section{Results}
	
This section presents the main results of the paper. Proofs are in the appendix.

 Theorem~\ref{thm:fsp} provides a finite-sample ATT bound for \emph{any} valid weight matrix from \textit{SCSA} (analogous to Theorems 1 and 2 of \textit{SCSA}) without the need for structural DGP assumptions. Note it expands on the DGPs considered in \textit{SCSA} and is
 therefore useful on its own for practitioners who wish to certify a
 finite-sample error bound for a pre-specified set of weights.
  
 Theorem~\ref{thm:consistency} shows consistency of \textit{SCSA} under the new assumptions with pre-specified weights. This theorem follows immediately from  Theorem \ref{thm:fsp}, however a different proof approach is used to emphasise that this result holds \emph{any} weighting
 method (conditional on weights)—classical DiD, separate SCM, pooled SCM, or machine-learned
 weights—as long as dispersion and donor restrictions hold.  
 
 Theorem~\ref{thm:scsa_consistency} shows consistency of \textit{SCSA} esimator with partial pooling and removing the pre-specified weights condition of Theorem~\ref{thm:consistency} allowing weights to be estimated by data following the procedure outlined in \textit{SCSA}. In order to gaurantee consistency uniform convergence of the imbalance functions is needed, which is shown in Lemma~\ref{lem:ULLN}.
 
 	Theorem~\ref{thm:scsa_consistency} is stated for $0<\nu<1$.
 The extreme choices $\nu\in\{0,1\}$ (Separate‐SCM and Pooled‐SCM) require
 additional homogeneity conditions (e.g., identical AR coefficients or factor
 loadings) to control the imbalance component that is not included in the
 objective.  Such DGP‐specific assumptions are avoided and therefore focus
 on the partially pooled estimator.

	\begin{thm}[Finite‑sample ATT error, arbitrary weights]
		\label{thm:fsp}
		Fix any $\Gamma\in\Delta_{\!scm}$ that satisfies the donor restriction
		$w_{ij}=0$ whenever $T_i\le T_j+k$ and the dispersion bound
		$\sum_i w_{ij}^{2}\le c<1$ (Assumption~\ref{assump:weights}b).
		Under Assumptions \ref{assump:anticip}--\ref{assump:subE}, for every
		$k\in\{0,\dots,K\}$,
		\[
		\Pr\!\Bigl(
		\bigl|\widehat{\ATT}_{k}-\ATT_{k}\bigr|
		\;>\;
		q_{\mathrm{pool}}(\Gamma)
		+L_{\min}^{-1/2}\,q_{\mathrm{sep}}(\Gamma)
		+ \sqrt{\tfrac{8 (1 + c) \log J}{cJ}} \, \kappa
		\Bigr)
		\;\le\;2J^{-2}.
		\]
	\end{thm}
	
	\paragraph{Remarks.}
	\begin{itemize}
		\item[(i)]  Because the bound depends on
		both $q_{\mathrm{pool}}$ and $q_{\mathrm{sep}}$,
		minimising a convex combination
		$\nu\,q_{\mathrm{pool}}^{2}+(1-\nu)\,q_{\mathrm{sep}}^{2}$ is a natural strategy; Theorem~\ref{thm:scsa_consistency} shows that the
		ridge-regularised minimiser \(\hat\Gamma\) indeed drives both terms to
		zero at a rate $\eta_{N,T}$, delivering consistency without structural
		DGP assumptions.

		\item[(ii)]  Equation \eqref{eq:3part} (see Appendix) separates the bias into a
		pooled component $q_{\mathrm{pool}}$ and a unit-specific component
		$q_{\mathrm{sep}}$, without any homogeneity or low-rank
		structure.  This decomposition is the formal motivation for the
		partially-pooled objective (weighting constant $\nu$) introduced in
		Lemma~\ref{lem:ULLN} and Theorem~\ref{thm:scsa_consistency}: driving either imbalance
		term to zero suffices for consistency.
		
		\item[(iii)]  
		Let $L_{\min}\!\to\!\infty$ and $J\!\to\!\infty$ while
		keeping  
		$q_{\mathrm{pool}}(\Gamma)\,\vee\,q_{\mathrm{sep}}(\Gamma)=o(1)$.  
		Because
		\[
		L_{\min}^{-1/2}\;\to\;0,
		\qquad
		\sqrt{\frac{\log J}{J}}\;\to\;0,
		\]
		the right–hand side of~\eqref{eq:3part}
		\[
		q_{\mathrm{pool}}(\Gamma)
		\;+\;
		L_{\min}^{-1/2}q_{\mathrm{sep}}(\Gamma)
		\;+\;
		\sqrt{\tfrac{8 (1 + c) \log J}{cJ}} \, \kappa
		\]
		converges to zero, hence
		$\bigl|\widehat{\operatorname{ATT}}_{k}-\operatorname{ATT}_{k}\bigr|
		=o_{p}(1)$.
		These are exactly the hypotheses of
Theorem~\ref{thm:consistency}, so the
		consistency result follows as an asymptotic corollary of
		Theorem~\ref{thm:fsp}.
		
	\end{itemize}
	
	The next example for Theorem \ref{thm:fsp} shows the importance of the parallel trends weights being able to close the gap between treated units and donors. 
	
	\begin{example}[Uniform weights and a non–vanishing pooled gap]
		\label{ex:biasfloor}
		
\mbox{}\\[.6\baselineskip]\noindent\textbf{Setup.}
		Let $J\ge1$ treated units and $N_0$ donors.  
		For every treated unit $j$ use \emph{uniform donor weights}
		$w_{ij}=1/N_0$.
		
		Assume that for each $j$ the treated pre-period mean exceeds
		the donor mean by the \emph{same} constant $\delta>0$:
		\[
		\bar Y^{0}_{j,\text{pre}}
		-\frac1{N_0}\sum_{i=1}^{N_0}\bar Y^{0}_{i,\text{pre}}
		\;=\;\delta .
		\]
		
		This homogeneity ensures the gaps do not average out across $j$.
		
		\medskip\noindent
		\textbf{Pooled imbalance.}  
		The homogeneity condition implies
		$q_{\mathrm{pool}}(\Gamma)=\delta$.
		
		\medskip\noindent
		\textbf{Separate imbalance.}  
		For each $j$, $q_j(\Gamma)=O(1)$ and therefore
		$q_{\mathrm{sep}}(\Gamma)=O(1)$.  
		Theorem~\ref{thm:fsp} multiplies this by $L_{\min}^{-1/2}$, so that
		term $\to0$ as $L_{\min}\to\infty$.
		
		\medskip\noindent
		\textbf{Finite-sample implication.}  
		Plugging $\Gamma$ into Theorem~\ref{thm:fsp} gives
		\[
		\Pr\!\Bigl(
		\bigl|\widehat{\operatorname{ATT}}_{k}-\operatorname{ATT}_{k}\bigr|
		> \delta
		+ \sqrt{\tfrac{8(1+c)\log J}{cJ}}\,\kappa
		\Bigr)
		\;\le\; 2J^{-2}.
		\]
		
	\end{example}

			Even as $L_{\min}\to\infty$ and $J\to\infty$, the deterministic bias
	floor $\delta$ remains; only the noise constant shrinks with $J$. This provides two insights on the importance of the weights.  If no weight matrix can achieve $q_{\mathrm{pool}}\!\to0$,
		Assumption~\ref{assump:C1} (convex-hull/spanning) fails—the donor
		pool simply cannot replicate the treated path. If better weights do exist, uniform weights are
		sub-optimal; the bias floor illustrates why estimating weights
		(Theorem~\ref{thm:scsa_consistency}) is essential. Either way, Theorem \ref{thm:fsp} shows that increasing $L$ or $J$ cannot rescue an estimator whose pooled imbalance is stuck at a positive constant.
	
	\begin{thm}[Convergence with pre-specified weights]\label{thm:consistency}
		Under
		Assumptions~\ref{assump:anticip}–\ref{assump:subE}, for every fixed event time \(k\ge0\),
		\begin{align*}
		\widehat{\ATT}_k \;&\xrightarrow{\;p\;}\; \ATT_k \text{ and} \\
		\bar{\widehat{\ATT}} &\xrightarrow{p} \bar{\ATT}.
	\end{align*}
	\end{thm}
	
\paragraph{Remarks.}
\begin{enumerate}
\item[(i)] As previously mentioned, this theorem follows immediately from Theorem~\ref{thm:fsp}. However, a different approach is used in the proof to show the result applies outside of \textit{SCSA} framework and only relying on weighted parallel trends instead of a structural DGP assumption.
\item[(ii)] Pointwise consistency relies on $J\to\infty$ to average away
post-treatment shocks; with fixed $J$ only the \emph{time-average}
$\bar\ATT$ is consistent as $K\to\infty$.
\end{enumerate}

\bigskip
\begin{lemma}[Uniform convergence of imbalance functionals]
	\label{lem:ULLN}
	Under Assumptions
	\ref{assump:anticip}, \ref{assump:sutva},
	\ref{assump:L},
	and \ref{assump:subE},
	\[
	\sup_{\Gamma\in\Delta_{\!scm}}
	\bigl|\widehat q_{\mathrm{pool}}^{\,2}(\Gamma)
	- q_{\mathrm{pool}}^{\,2}(\Gamma)\bigr|
	\;\xrightarrow{p}\;0,
	\quad
	\sup_{\Gamma\in\Delta_{\!scm}}
	\bigl|\widehat q_{\mathrm{sep}}^{\,2}(\Gamma)
	- q_{\mathrm{sep}}^{\,2}(\Gamma)\bigr|
	\;\xrightarrow{p}\;0 .
	\]
\end{lemma}

		\paragraph{Remarks.}
\begin{enumerate}
	\item[(i)] Uniform convergence ensures the
	sample‐chosen weights behave as if the population criteria were observed.
	Without it, the arg-min step in Theorem \ref{thm:scsa_consistency} could drift away from the
	population minimiser.
	\item[(ii)] The lemma plays the same technical role as a Glivenko–Cantelli (uniform LLN) result: it guarantees that the empirical imbalance criteria converge uniformly to their population analogues. The novelty here is to establish the result for the staggered-adoption imbalance class under only sub-exponential tails, with no factor-model or AR structure. 
	\item[(iii)] The covering number in Lemma \ref{lem:ULLN} uses $L_{\max}$ because the empirical mean that defines $\widehat q_{\text{pool}}$ averages over $\ell=1,\dots,L_{\max}$.
	\item[(iv)] The error is shown to decay at rate $L_{\max}^{-1/3}$. A faster rate of $L_{\max}^{-1/2}$ could be attained under sub-Gaussian tails.
	\item[(v)] The covering number condition requires $(N_0-1)J=o(L_{\max}/\log L_{\max})$; hence
	Lemma~\ref{lem:ULLN} holds even if $J$ is fixed, provided $L_{\max}$ grows
	sufficiently fast.
\end{enumerate}

	\vspace{.5em}
	
	\begin{thm}[Consistency with estimated \textit{SCSA} weights]
		\label{thm:scsa_consistency}
		Adopt Assumptions
		\ref{assump:anticip}--\ref{assump:C2}.
		Define the data–driven weight matrix
		\[
		\hat\Gamma
		\;=\;
		\argmin_{\Gamma\in\Delta_{\!scm}}
		\Bigl\{
		\nu\,\widehat q_{\mathrm{pool}}^{\,2}(\Gamma)
		+(1-\nu)\,\widehat q_{\mathrm{sep}}^{\,2}(\Gamma)
		+\lambda\|\Gamma\|_{F}^{2}
		\Bigr\},
		\qquad 0<\nu<1 .
		\]
		Form the intercept-shifted unit–level and average estimators
		\(\hat\tau_{jk}\) and \(\widehat{\ATT}_k\) with these weights.
		Then, for every fixed event time \(k\ge0\),
		\begin{align*}
	\widehat{\ATT}_k \;&\xrightarrow{\;p\;}\; \ATT_k \text{ and} \\
	\bar{\widehat{\ATT}} &\xrightarrow{p} \bar{\ATT}.
\end{align*}
	\end{thm}
	
		\paragraph{Remarks.}
		\begin{enumerate}
			\item[(i)] The theorem covers any
			ridge‑regularised partially pooled SCM, including the popular
			\textsf{augsynth} reccomended hueristic $\nu$  or $\nu=1/2$, as well as
			hyper‑parameter choices selected by cross‑validation \citep{augsynth}.
			\item[(ii)] Consistency under pure separate‑SCM ($\nu=0$) or pure pooled‑SCM ($\nu=1$) requires extra homogeneity (all AR coefficients equal, or constant factor loadings) because one of the two imbalance components is left uncontrolled. Thus $0<\nu<1$ in routine applications is reccomended.
			\item[(iii)] The plug‑in bound derived from Theorem \ref{thm:fsp} suggests choosing $\lambda$ of the same order as the empirical imbalance squared; this informal rule balances bias and variance in moderate samples.
			\item[(iv)] When $J$ is fixed, the variance term
			$\sqrt{\log J/J}$ in the finite-sample bound does not vanish;
			increasing $L_{\max}$ circumvents this by averaging over an expanding post-treatment
			window $K$.
		\end{enumerate}

		\section{Practical diagnostics}
	Pre-treatment placebo gaps remain the most convenient empirical
	check.  Small root-mean-squared gaps suggest Assumption~
	\ref{assump:wpt} is plausible.\footnote{%
		See \citet{bilinski2020herenoninferiorityapproachesparallel} for a discussion of presenting evidence about parallel trends in a non-inferiority framework.}
	Weight dispersion can be summarised by the effective-donor count
	\(m_j=1/\sum_i w_{ij}^2\); a rule-of-thumb threshold \(m_j\ge4\)
	helps ensure Assumption~\ref{assump:weights}b.  Finally, analysts should report sensitivity and placebo gap RMSE to shortening or lengthening the pre‑treatment window and/or inclusion of addition donors; analysts may also examine robustness to smooth deviations from parallel trends using the  \cite{rambachan2023more} framework.

	\section{Conclusion}
	
	This paper broadens the theoretical foundations of augmented
	synthetic–control estimators in staggered–adoption settings.  By
	replacing factor or autoregressive data–generating assumptions with the
	design–based \emph{weighted parallel-trends} condition and mild
	sub-exponential tails, in this paper a (i) finite-sample error bound that
	separates pooled and unit–specific imbalance is derived, (ii) consistency
	for any pre-specified weighting scheme satisfying simple
	dispersion constraints is established, and (iii) an arg-min consistency result for
	the data-driven, partially–pooled ridge estimator of Ben-Michael,
	Feller and Rothstein (2022) is provided.  The finite-sample bound implies that, in principle, driving either imbalance component to zero suffices for consistency; the partially pooled objective ($0<\nu<1$) provides a practical, model-free way to do so by controlling both components simultaneously.
	
	From an applied perspective the new conditions map directly onto
	familiar diagnostics.  Root-mean-squared placebo gaps test weighted
	parallel trends; the effective-donor count and ridge path trace weight
	dispersion; window-length and donor-set sensitivity illuminate the
	convex-hull requirement.  Because all results hinge only on these
	observable quantities, researchers can help ensure assumptions without
	taking a stand a structural DGP.
	
	Several extensions are immediate.  Allowing auxiliary pretreatment
	covariates would bring the theory fully in line with \textsf{augsynth};
	bootstrap-based uncertainty quantification could adapt the finite-sample
	bound; and a systematic simulation study would benchmark finite-sample
	performance against recent alternatives such as synthetic DiD. These directions are left for future work and hope the present results encourage
	broader use of intercept-augmented SCM in policy evaluation.

	\newpage
	
	\appendix
	\section{Proofs}
	
	\subsection{Theorem \ref{thm:fsp}}

	\begin{proof}
		For each treated unit $j$ and lag $\ell\ge0$ define
		\[
		Z^{(\Gamma)}_{j\ell}
		:= Y^{0}_{j,T_j-\ell}
		-\sum_{i}w_{ij}Y^{0}_{i,T_j-\ell}.
		\]
		Let
		
		\begin{align*}
			q_{\mathrm{pool}}^{2}(\Gamma)
			&= \frac1L_{\max} \sum_{\ell=1}^{L_{\max}}
			\Bigl\{ \frac1J
			\sum_{j: T_{j} > \ell} Z^{(\Gamma)}_{j\ell}
			\Bigr\}^{2}, \\[4pt]
			q_{j}^{2}(\Gamma)
			&= \frac1{L_{j}} \sum_{\ell=1}^{L_{j}}
			\bigl[ Z^{(\Gamma)}_{j\ell} \bigr]^{2}, \text{and} \\[4pt]
			q_{\mathrm{sep}}^{2}(\Gamma)
			&= \frac1J \sum_{j=1}^{J} q_{j}^{2}(\Gamma).
		\end{align*}
		
		\paragraph{Step 1 (ATT estimator in residual form).}
		The intercept-shifted estimator for unit $j$
		(\(k\) periods after adoption) is
		\[
		\hat\tau_{jk}
		=\! \Bigl[ Y_{j,T_j+k}
		-\!\sum_{i}w_{ij}Y_{i,T_j+k}\Bigr]
		-\Bigl[\bar Y_{j,\mathrm{pre}}
		-\!\sum_{i}w_{ij}\bar Y_{i,\mathrm{pre}}\Bigr].
		\]
		It follows that
		\begin{align*}
			\hat\tau_{jk}-\tau_{jk}
			&=   Y^{0}_{j,T_j+k}
			-\sum_{i}w_{ij}Y^{0}_{i,T_j+k}
			-\bar Y^{0}_{j,\mathrm{pre}}
			+\sum_{i}w_{ij}\bar Y^{0}_{i,\mathrm{pre}} \\
			&=   Z^{(\Gamma)}_{j,-k}
			-\bar Z^{(\Gamma)}_{j,\mathrm{pre}}
			+\varepsilon_{j,T_j+k}
			-\sum_{i}w_{ij}\varepsilon_{i,T_j+k},
		\end{align*}
		where $Z^{(\Gamma)}_{j,-k}$ abbreviates the
		\emph{forward} residual
		$Y^{0}_{j,T_j+k}-\sum_i w_{ij}Y^{0}_{i,T_j+k}$.
		
		Then
		\begin{equation}\label{eq:3part}
			\widehat{\ATT}_{k}-\ATT_{k}
			\;=\;
			\underbrace{\frac1J\sum_{j}Z^{(\Gamma)}_{j,-k}}
			_{\text{pooled imbalance}}
			\;-\;
			\underbrace{\frac1J\sum_{j}\bar Z^{(\Gamma)}_{j,\mathrm{pre}}}
			_{\text{separate imbalance}}
			\;+\;
			\underbrace{\frac1J\sum_{j}
				\bigl(\varepsilon_{j,T_j+k}
				-\sum_i w_{ij}\varepsilon_{i,T_j+k}\bigr)}
			_{\text{post-treatment noise}}.
		\end{equation}
		
		\paragraph{Step 2  (Deterministic bounds for the imbalance terms).}
		By Cauchy–Schwarz, the pooled term is bounded by
		\[
		\bigl|\tfrac1J\sum_j Z^{(\Gamma)}_{j,-k}\bigr|
		\le q_{\mathrm{pool}}(\Gamma)
		\]
		and the separate term is bounded by
		\[
		\Bigl|\tfrac1J\sum_j \bar Z^{(\Gamma)}_{j,\mathrm{pre}}\Bigr|
		\le L_{\min}^{-1/2}\Bigl\{\tfrac1J
		\sum_j q_{j}^{2}(\Gamma)\Bigr\}^{1/2}
		=  L_{\min}^{-1/2}\,q_{\mathrm{sep}}(\Gamma).
		\]

		\paragraph{Step 3  (Concentration for the noise term).}
		For each treated unit $j$ let
		\(
		\xi_{j}:=\varepsilon_{j,T_j+k}-\sum_i w_{ij}\varepsilon_{i,T_j+k}.
		\)
		Because $\|\xi_{j}\|_{\psi_{1}}\le 2\kappa\sqrt{1+c},$
		Proposition 2.7.1 of \citet{vershynin2018high} yields
		
		\[\Pr\!\Bigl(
		\bigl|\tfrac1J\sum_{j=1}^{J}\xi_{j}\bigr|
		> C_{3}\sqrt{\tfrac{\log J}{J}}
		\Bigr)
		\;\le\; 2J^{-2},
		\quad
		C_{3}:=\sqrt{\tfrac{8}{c}}\kappa\sqrt{1+c}.\]
		
		Hence, with probability at least $1-2J^{-2}$,
		\[
		\Bigl|\tfrac1J\sum_{j}\xi_{j}\Bigr|
		\;\le\; C_{3}\sqrt{\tfrac{\log J}{J}}.
		\]
		
		\paragraph{Step 4 (Assemble the pieces).}
		The equation \eqref{eq:3part}, the deterministic bounds (Step 2) and the
		probabilistic bound (Step 3) together imply the event
		\[
		\bigl|\widehat{\ATT}_{k}-\ATT_{k}\bigr|
		\;\le\;
		q_{\mathrm{pool}}(\Gamma)
		+ L_{\min}^{-1/2}\,q_{\mathrm{sep}}(\Gamma)
		+ \sqrt{\tfrac{8 (1 + c) \log J}{cJ}} \, \kappa
		\]
		
		fails with probability at most $2J^{-2}$.
	\end{proof}

	\subsection{Theorem \ref{thm:consistency}}

	\begin{proof}
		Consistency is proven in two steps: unbiasedness and diminishing variance. 
		
		\paragraph{Step 1 (Unbiasedness).}
		Define \(Z_{it}^0:=Y_{it}^0-\bar Y_{i,\text{pre}}\). For any $k\ge0$,
		\[
		\mu_{j,T_j+k}-\sum_i w_{ij}\mu_{i,T_j+k}
		\;=\;\sum_{t=1}^{k}\Bigl(
		\Delta\mu_{j,T_j+t}
		-\sum_i w_{ij}\Delta\mu_{i,T_j+t}\Bigr).
		\]
		Using SUTVA (Assumptions~\ref{assump:anticip}–\ref{assump:sutva}) and by Assumption~\ref{assump:wpt} each summand has mean 0 for $t\le0$,
		so in expectation only
		$\E[\tau_{jk}]$ remains:

		\[
		\E[\hat\tau_{jk}]
		=\E[\Delta Z_{j,T_j+k}^0]
		-\sum_i w_{ij}\E[\Delta Z_{i,T_j+k}^0]
		=0+\E[\tau_{jk}].
		\]
		Hence \(\E[\widehat{\ATT}_k]=\ATT_k\).
		
		\paragraph{Step 2 (Diminishing variance).}
		Write
		\[
		\hat\tau_{jk}-\tau_{jk}
		=(\varepsilon_{j,T_j+k}-\sum_i w_{ij}\varepsilon_{i,T_j+k})
		-(\bar\varepsilon_{j,\text{pre}}
		-\sum_i w_{ij}\bar\varepsilon_{i,\text{pre}}).
		\]
		Sub-exponentiality plus independence gives
		\(
		\Var(\hat\tau_{jk})
		= \kappa^2\bigl(1+\sum_i w_{ij}^2
		+ L_j^{-1}+\sum_i w_{ij}^2 L_i^{-1}\bigr)
		\le C(\sum_i w_{ij}^2 + L_{\min}^{-1}),
		\)
		for a constant \(C\).
		Assumptions~\ref{assump:weights}b and~\ref{assump:L} imply
		\(\sup_j\Var(\hat\tau_{jk})\le C^\ast<\infty\) and
		\(L_{\min}^{-1}\to0\).
		Because the \(\hat\tau_{jk}\) are independent across \(j\),
		\(
		\Var(\widehat{\ATT}_k)
		=J^{-2}\sum_{j}\Var(\hat\tau_{jk})
		\le C^\ast/J \to 0.
		\)
		
		Unbiasedness plus vanishing variance yields
		$\widehat{\ATT}_k \xrightarrow{p} \ATT_k$ 
		whenever $J\to\infty$, and
		$\bar{\widehat{\ATT}}\xrightarrow{p}\bar{\ATT}$ 
		whenever $J\,K\to\infty$.
		
	\end{proof}

	\subsection{Lemma \ref{lem:ULLN}}

	\paragraph{Sketch of proof.}
	For any fixed \(\Gamma\) the centred average
	\(L^{-1}\sum_{\ell=1}^{L}\bigl[(\zeta_\ell^{(\Gamma)})^{2}
	-\mathbb E(\zeta_\ell^{(\Gamma)})^{2}\bigr]\)
	is the mean of \(L_{\max}\) independent sub-exponential
	random variables.  Bernstein’s inequality therefore yields an
	\(e^{-cL_{\max}t^{2}}\) tail.  A covering argument shows that
	\(\Delta_{\!scm}\subset\mathbb R^{N_{0}J}\) admits an
	\(\ell_{1}\)-\(\varepsilon\)-net of cardinality
	\((C/\varepsilon)^{N_{0}J}\);  
	taking \(\varepsilon_{L_{\max}}=L_{\max}^{-2}\) makes
	\(\log|\mathcal N_{\varepsilon_L}|=o(L_{\max})\).
	A union bound over the net plus a Lipschitz extension to the whole
	simplex completes the proof.
	\hfill\(\triangleleft\)
	
	\begin{proof}
		The pooled criterion is shown; the separate version repeats the
		same steps unit-by-unit.
		
		\medskip
		\paragraph{Step 1 (Empirical-process form).}
		For a fixed weight matrix \(\Gamma=(\gamma_{ij})\in\Delta_{\!scm}\) define
		\[
		\zeta^{(\Gamma)}_{\ell}
		:=\frac1J\sum_{j:T_{j}>\ell}
		\Bigl(
		Y^{0}_{j,T_{j}-\ell}
		-\sum_{i}\gamma_{ij}Y^{0}_{i,T_{j}-\ell}
		\Bigr),
		\qquad \ell=1,\dots,L_{\max} .
		\]
		By construction
		\(
		\widehat q_{\mathrm{pool}}^{\,2}(\Gamma)
		=L_{\max}^{-1}\sum_{\ell=1}^{L_{\max}}(\zeta^{(\Gamma)}_{\ell})^{2},
		\)
		while (no-anticipation and stationarity in \(\ell\))
		\(
		q_{\mathrm{pool}}^{\,2}(\Gamma)
		=\mathbb E\!\bigl[(\zeta^{(\Gamma)}_{1})^{2}\bigr].
		\)
		
		\medskip
		\paragraph{Step 2 (Single-\(\Gamma\) concentration).}
		Write
		\(Y^{0}_{it}=\mu_{it}+\varepsilon_{it}\)
		with independent \(\kappa\)-sub-exponential
		\(\varepsilon_{it}\) (Assumption \ref{assump:subE}).
		Because
		\(
		\zeta^{(\Gamma)}_{\ell}-\mathbb E\zeta^{(\Gamma)}_{\ell}
		=\sum_{i,j}a^{(\Gamma)}_{ij}\varepsilon_{it}
		\)
		for suitable coefficients \(a^{(\Gamma)}_{ij}\)
		(independent across \(\ell\)), then
		\[
		\bigl\|\,\zeta^{(\Gamma)}_{\ell}
		-\mathbb E\zeta^{(\Gamma)}_{\ell}\bigr\|_{\psi_{1}}
		\;\le\;2\kappa
		\sqrt{1+\sum_{ij}\gamma_{ij}^{2}}
		\;\le\;2\kappa\sqrt{1+c},
		\]
		by \cite{vershynin2018high} Proposition 2.7.1 and Assumption \ref{assump:weights}.
		Each centred \(\ell\)-summand
		\(
		X^{(\Gamma)}_{\ell}
		:=(\zeta^{(\Gamma)}_{\ell})^{2}
		-\mathbb E(\zeta^{(\Gamma)}_{\ell})^{2}
		\)
		is therefore sub-Weibull   \citep{vladimirovasubweibull2020} with bound $K=2 \kappa\sqrt{1+c}\).
		Bernstein’s inequality for the \emph{mean} of \(L_{\max}\) such variables gives,
		for every \(t>0\),
		\begin{equation}\label{eq:Bern}
			\Pr\!\Bigl(
			\bigl|\widehat q_{\mathrm{pool}}^{\,2}(\Gamma)
			-q_{\mathrm{pool}}^{\,2}(\Gamma)\bigr|>t
			\Bigr)
			\;\le\;
			2\exp\!\Bigl(
			-\frac{L_{\max}\,t^{2}}{32\kappa^{2}(1+c)}
			\Bigr).
		\end{equation}
		
		\medskip
		\paragraph{Step 3 (An $\boldsymbol{\varepsilon}$-net for the
			simplex).}
		The feasible set is a Cartesian product of \(J\) probability
		simplices in \(\mathbb R^{N_{0}}\), an affine set of dimension
		\(d=(N_{0}-1)J\).
		By Pollard’s covering bound
		\citep[Lemma 4.1]{pollard1990empirical},
		\[
		N\bigl(\varepsilon,\Delta_{\!scm},\lVert\cdot\rVert_{1}\bigr)
		\;\le\;\Bigl(\tfrac{3}{\varepsilon}\Bigr)^{d},
		\qquad 0<\varepsilon<1.
		\]
		Choose \(\varepsilon_{L}=L_{\max}^{-2}\) and let
		\(\mathcal N_{\varepsilon_{L_{\max}}}\) be a minimal $\boldsymbol{\varepsilon}$-net.
		Then
		\[
		\log|\mathcal N_{\varepsilon_{L_{\max}}}|
		\;\le\;d\log(3L_{\max}^{2})
		\;=\;
		(N_{0}J)\,\bigl(2\log L_{\max}+\log 3\bigr)
		\;=\;o(L_{\max})
		\]
		provided \(N_{0}J=o(L_{\max}/\log L_{\max})\); polynomial growth in \(L_{\max}\) with total
		exponent \(<1\) meets this condition, as allowed by
		Assumption \ref{assump:L}.
		
		\medskip
		\paragraph{Step 4 (Union bound over the net).}
		From \eqref{eq:Bern},
		\[
		\Pr\!\Bigl(
		\max_{\Gamma\in\mathcal N_{\varepsilon_{L_{\max}}}}
		\bigl|\widehat q_{\mathrm{pool}}^{\,2}(\Gamma)
		-q_{\mathrm{pool}}^{\,2}(\Gamma)\bigr|
		>t
		\Bigr)
		\;\le\;
		2|\mathcal N_{\varepsilon_{L_{\max}}}|
		\exp\!\Bigl\{
		-L_{\max}\,t^{2}\bigl/ \!\bigl[32\kappa^{2}(1+c)\bigr]
		\Bigr\}.
		\]
		Because \(\log|\mathcal N_{\varepsilon_{L_{\max}}}|=o(L_{\max})\),
		choose any \(t_{L_{\max}}\downarrow0\) with \(Lt_{L_{\max}}^{2}\to\infty\), e.g.
		\(t_{L_{\max}}=L_{\max}^{-1/3}\), and the probability above tends to zero.
		
		\medskip
		\medskip
		\paragraph{Step 5 (Extend from the $\varepsilon$–net to all of
			$\boldsymbol{\Delta_{\!scm}}$).}
		Let $\Gamma\in\Delta_{\!scm}$ be arbitrary and choose
		$\tilde\Gamma\in\mathcal N_{\varepsilon_{L_{\max}}}$ with
		$\lVert\Gamma-\tilde\Gamma\rVert_1\le\varepsilon_{L_{\max}}=L_{\max}^{-2}$.
		Write
		\[
		\delta_{ij}:=\gamma_{ij}-\tilde\gamma_{ij},
		\qquad  
		D:=\lVert\Gamma-\tilde\Gamma\rVert_1
		=\sum_{ij}|\delta_{ij}|\;(=\varepsilon_{L_{\max}}).
		\]
		
		\paragraph{Step 5a (A high-probability stochastic envelope).}
		Because $\mu_{it}$ is bounded and the shocks
		$\varepsilon_{it}$ are $\kappa$–sub-exponential,
		a union bound over the $(N_0+J)T$ observations gives
		\begin{equation}\label{eq:maxY}
			B_T \;:=\; C\sqrt{\log\!\bigl((N_0+J)T\bigr)}
			\quad\Longrightarrow\quad
			\Pr\!\Bigl(\max_{i,t}|Y^{0}_{it}|\le B_T\Bigr)\;=\;1-o(1),
		\end{equation}
		for some absolute constant $C>1$.  (Take
		$C\ge\kappa\sqrt{32(1+c)}$ to dominate the sub-exponential tail.)
		
		\paragraph{Step 5b (Lipschitz continuity in $\ell_1$ norm).}
		Abbreviate
		$Z^{(\Gamma)}_{j\ell}=Y^{0}_{j,T_j-\ell}
		-\sum_{i}\gamma_{ij}Y^{0}_{i,T_j-\ell}$.
		Then
		\[
		\widehat q_{\mathrm{pool}}^{\,2}(\Gamma)
		=\frac1{JL_{\max}}\sum_{\ell=1}^{L_{\max}}
		\sum_{j:T_j>\ell} \bigl[Z^{(\Gamma)}_{j\ell}\bigr]^2 .
		\]
		On the event \(\left\{\displaystyle\max_{i,t}|Y^{0}_{it}|\le B_T\right\}\) then
		\[
		|Z^{(\Gamma)}_{j\ell}|
		\;\le\;
		2 B_T ,\qquad
		|Z^{(\Gamma)}_{j\ell}-Z^{(\tilde\Gamma)}_{j\ell}|
		\;\le\;
		B_T \sum_{i}|\delta_{ij}|
		\;\le\; B_T D .
		\]
		Hence, using the inequality
		$|x^{2}-y^{2}|\le(|x|+|y|)\,|x-y|$,
		\[
		\bigl|\widehat q_{\mathrm{pool}}^{\,2}(\Gamma)
		-\widehat q_{\mathrm{pool}}^{\,2}(\tilde\Gamma)\bigr|
		\;\le\;
		\frac{1}{JL}\sum_{\ell=1}^{L}\sum_{j:T_j>\ell}
		4 B_T \,(B_T D)
		\;\le\;
		\frac{4B_T^{2}}{J}\,\varepsilon_{L_{\max}} .
		\tag{C}
		\]

		The same calculation (with expectations) shows that
		\(
		\bigl|q_{\mathrm{pool}}^{\,2}(\Gamma)
		-q_{\mathrm{pool}}^{\,2}(\tilde\Gamma)\bigr|
		\le 4B_T^{2}J^{-1}\varepsilon_{L_{\max}}
		\)
		on the same high-probability set.
		
		\paragraph{Step 5c (Radius term vanishes).}
		Because $B_T^{2}=O(\log T)$ by (\ref{eq:maxY}) and
		$\varepsilon_{L_{\max}}=L_{\max}^{-2}$, while
		$L_{\min}\to\infty$ by Assumption~\ref{assump:L},
		then
		\[
		\frac{4B_T^{2}}{J}\,\varepsilon_{L_{\max}}
		=O\!\bigl(L_{\max}^{-2}\log T\bigr)
		\;\xrightarrow{p}\;0 .
		\]
		Consequently, the supremum deviation proved over the
		$\varepsilon_{L_{\max}}$–net in Step~4 transfers to the entire simplex.
		
		\medskip
		Combining Steps 4 and 5 shows
		\[
		\sup_{\Gamma\in\Delta_{\!scm}}
		\bigl|
		\widehat q_{\mathrm{pool}}^{\,2}(\Gamma)
		-q_{\mathrm{pool}}^{\,2}(\Gamma)
		\bigr|
		\;=\;o_{p}(1).
		\]
		Repeating the identical argument for each treated unit \(j\) proves the
		result for \(\widehat q_{\mathrm{sep}}^{\,2}\) and completes the proof.
	\end{proof}
	
	\subsection{Theorem \ref{thm:scsa_consistency}}

	\begin{proof}
		\emph{Notation.}  Write
		\[
		q_{\mathrm{pool}}^{2}(\Gamma)
		:=\E[\zeta_{1}^{(\Gamma)}]^{2},
		\quad
		q_{\mathrm{sep}}^{2}(\Gamma)
		:=\frac1J\sum_{j=1}^{J}
		\E\bigl[\zeta_{j,1}^{(\Gamma)}\bigr]^{2},
		\]
		and let \( \widehat q_{\mathrm{pool}}^{2},\widehat q_{\mathrm{sep}}^{2} \) be the
		corresponding sample versions.
		Define
		\[
		\Psi_{N,T}(\Gamma)
		:=\nu\,q_{\mathrm{pool}}^{2}(\Gamma)
		+(1-\nu)\,q_{\mathrm{sep}}^{2}(\Gamma)
		+\lambda\|\Gamma\|_{F}^{2},
		\]
		with $\widehat\Psi_{N,T}(\Gamma)$ the sample analog.
		
		\paragraph{Step 1 (Uniform law of large numbers and arg-min	consistency).}
		Lemma~\ref{lem:ULLN} gives a uniform LLN:
		\[
		\sup_{\Gamma\in\Delta_{\!scm}}
		\bigl|\widehat\Psi_{N,T}(\Gamma)-\Psi_{N,T}(\Gamma)\bigr|
		\;=\;o_{p}(1).
		\tag{1}
		\]
		Assumption~\ref{assump:C1} guarantees that there exists some
		\(\Gamma^{\star}\in\Delta_{\!scm}\) with imbalance
		\(q_{\mathrm{pool}}(\Gamma^{\star})\vee q_{\mathrm{sep}}(\Gamma^{\star})
		\le\eta_{N,T}\)
		and Frobenius norm bounded by \(c^{1/2}\); hence
		\[
		\Psi_{N,T}(\Gamma^{\star})
		\;\le\;
		\nu\,\eta_{N,T}^{2}+(1-\nu)\,\eta_{N,T}^{2}
		+\lambda c
		\;=\;O(\eta_{N,T}^{2}).
		\tag{2}
		\]
		Combine (1)--(2) and apply the Arg-min Theorem
		\citep[Theorem~5.7]{van1998asymptotic} to the objective
		\(\widehat\Psi_{N,T}\):
		\[
		\widehat\Psi_{N,T}(\hat\Gamma)
		\;\le\;
		\widehat\Psi_{N,T}(\Gamma^{\star})
		= \Psi_{N,T}(\Gamma^{\star})+o_{p}(1)
		= O_{p}(\eta_{N,T}^{2}),
		\]
		so
		\[
		q_{\mathrm{pool}}\!\bigl(\hat\Gamma\bigr)=O_{p}(\eta_{N,T}),
		\qquad
		q_{\mathrm{sep}}\!\bigl(\hat\Gamma\bigr)=O_{p}(\eta_{N,T}).
		\tag{A}
		\]
		
		\paragraph{Step 2 (Controlling the ridge penalty).}
		Because $\Psi_{N,T}(\hat\Gamma)\ge\lambda\|\hat\Gamma\|_{F}^{2}$,
		\[
		\lambda\|\hat\Gamma\|_{F}^{2}
		\;\le\;O_{p}(\eta_{N,T}^{2}),
		\qquad\Longrightarrow\qquad
		\|\hat\Gamma\|_{F}^{2}=O_{p}\bigl(\eta_{N,T}^{2}/\lambda\bigr).
		\]
		Assumption~\ref{assump:C2} imposes $\eta_{N,T}^{2}/\lambda\to0$,
		hence $\|\hat\Gamma\|_{F}^{2}=O_{p}(1)$, so the dispersion bound
		holds w.p. $\!1-o(1)$.

		\paragraph{Step 3 (Error bound for the estimated SCM).}
		Plug \(\Gamma=\hat\Gamma\) into Theorem~\ref{thm:fsp}.  The pooled and separate
		imbalances are \(O_{p}(\eta_{N,T})\) by (A), while the Frobenius norm is
		\(O_{p}(1)\) by Step~2.  Theorem~\ref{thm:fsp} therefore yields
		
			\[
		\bigl|\widehat{\ATT}_k-\ATT_k\bigr|
		\;\le\;
		C_{1}\,\eta_{N,T}
		\;+\;
		C_{2}\sqrt{\tfrac{\log J}{J}}
		\;+\;
		o_{p}(1),
		\tag{3}
		\]
		for constants \(C_{1},C_{2}\) that do not depend on \(N,T\).
		
		\paragraph{Step 4 (Conclude consistency).}

		Because $\eta_{N,T}\to0$ and $\sqrt{\log J/J}\to0$ when $J\to\infty$, the
		right-hand side of \eqref{eq:3part} converges to $0$, proving
		$\widehat{\ATT}_k\xrightarrow{p}\ATT_k$.
		If instead $J$ is fixed while $K\to\infty$, the identical bound with
		variance $1/(J\,K)$ shows
		$\bar{\widehat{\ATT}}\xrightarrow{p}\bar{\ATT}$.
		
	\end{proof}

	\vspace{1em}
	\bibliographystyle{apalike}
	\bibliography{scm-consistency}

\begin{thebibliography}{}

\bibitem[Abadie et~al., 2010]{abadie2010synthetic}
Abadie, A., Diamond, A., and Hainmueller, J. (2010).
\newblock Synthetic control methods for comparative case studies: Estimating
  the effect of {C}alifornia’s tobacco control program.
\newblock {\em Journal of the American Statistical Association},
  105(490):493--505.

\bibitem[Abadie et~al., 2015]{abadie2015comparative}
Abadie, A., Diamond, A., and Hainmueller, J. (2015).
\newblock Comparative politics and the synthetic control method.
\newblock {\em American Journal of Political Science}, 59(2):495--510.

\bibitem[Arkhangelsky et~al., 2021]{arkhangelsky2021synthetic}
Arkhangelsky, D., Athey, S., Hirshberg, D.~A., Imbens, G.~W., and Wager, S.
  (2021).
\newblock Synthetic difference-in-differences.
\newblock {\em American Economic Review}, 111(12):4088--4118.

\bibitem[Athey and Imbens, 2021]{athey2021design}
Athey, S. and Imbens, G.~W. (2021).
\newblock Design-based analysis in difference-in-differences settings with
  staggered adoption.
\newblock {\em Journal of Econometrics}, 225(2):1--17.

\bibitem[Ben{-}Michael, 2025]{augsynth}
Ben{-}Michael, E. (2025).
\newblock augsynth: Augmented synthetic control method.
\newblock https://github.com/ebenmichael/augsynth.
\newblock Accessed: 2025-05-15.

\bibitem[Ben{-}Michael et~al., 2022]{benmichael2022synthetic}
Ben{-}Michael, E., Feller, A., and Rothstein, J. (2022).
\newblock Synthetic controls with staggered adoption.
\newblock {\em Journal of the Royal Statistical Society: Series B},
  84(2):351--381.

\bibitem[Bilinski and Hatfield,
  2020]{bilinski2020herenoninferiorityapproachesparallel}
Bilinski, A. and Hatfield, L.~A. (2020).
\newblock Nothing to see here? {N}on-inferiority approaches to parallel trends
  and other model assumptions.

\bibitem[Callaway and Sant'Anna, 2021]{callaway2021difference}
Callaway, B. and Sant'Anna, P. H.~C. (2021).
\newblock Difference-in-differences with multiple time periods.
\newblock {\em Journal of Econometrics}, 225(2):200--230.

\bibitem[Doudchenko and Imbens, 2016]{doudchenko2016difference}
Doudchenko, N. and Imbens, G.~W. (2016).
\newblock Balancing, regression, difference-in-differences and synthetic
  control methods: A synthesis.
\newblock (22791).

\bibitem[Pollard, 1990]{pollard1990empirical}
Pollard, D. (1990).
\newblock {\em Empirical processes : theory and applications}.
\newblock Institute of Mathematical Statistics.

\bibitem[Rambachan and Roth, 2023]{rambachan2023more}
Rambachan, A. and Roth, J. (2023).
\newblock A more credible approach to parallel trends.
\newblock {\em Review of Economic Studies}, 90(1):58--94.

\bibitem[Stefano and Mellace, 2024]{inclusiveSCM2024}
Stefano, R.~D. and Mellace, G. (2024).
\newblock The inclusive synthetic control method.

\bibitem[Sun and Abraham, 2021]{sun2021estimating}
Sun, L. and Abraham, S. (2021).
\newblock Estimating dynamic treatment effects in event studies with
  heterogeneous treatment effects.
\newblock {\em Journal of Econometrics}, 225(2):175--199.

\bibitem[van~der Vaart, 1998]{van1998asymptotic}
van~der Vaart, A.~W. (1998).
\newblock {\em Asymptotic Statistics}.
\newblock Cambridge University Press.

\bibitem[Vershynin, 2018]{vershynin2018high}
Vershynin, R. (2018).
\newblock {\em High-Dimensional Probability: An Introduction with Applications
  in Data Science}.
\newblock Cambridge Series in Statistical and Probabilistic Mathematics.
  Cambridge University Press, Cambridge, UK.

\bibitem[Vladimirova et~al., 2020]{vladimirovasubweibull2020}
Vladimirova, M., Girard, S., Nguyen, H., and Arbel, J. (2020).
\newblock Sub-weibull distributions: Generalizing sub-gaussian and
  sub-exponential properties to heavier tailed distributions.
\newblock {\em Stat}, 9(1):e318.
\newblock e318 sta4.318.

\end{thebibliography}
\end{document}